\documentclass[12pt]{article}
\usepackage{epsfig}
\usepackage{amsmath,amsfonts}
\usepackage[square,numbers]{natbib}
\usepackage{amssymb,graphicx}

\def\pa{\partial}
\def\L{\Lambda}
\def\s{\sigma}

\def\k{\kappa}

\def\d{\delta}
\def\r{\rho}
\def\h{\hat}

\def\l{\label}

\def\la{\lambda}
\def\g{\gamma}
\def\m{\mu}
\def\n{\nu}
\def\e{\epsilon}
\def\be{\begin{equation}}
\def\ee{\end{equation}}
\def\ba{\begin{eqnarray}}
\def\ea{\end{eqnarray}}

\def\h{\hat}
\def\b{\bar}
\def\t{\tilde}
\def\f{\frac}
\def\G{\Gamma}

\def\e{\epsilon}

\begin{document}

\begin{center}{\Large\bf
Energy-momentum tensor of bouncing gravitons }
\end{center}
\vspace{1cm}
\centerline{\bf Mikhail Z. Iofa \footnote{ e-mail:
iofa@theory.sinp.msu.ru}} \centerline{Skobeltsyn Institute of Nuclear
Physics}
\centerline{Moscow State University}
\centerline{Moscow 119992, Russia}
\vspace{1 cm}

\begin{abstract}

In models of the Universe with extra dimensions gravity propagates in 
the whole space-time. Graviton production by matter on the brane is 
significant in the early hot Universe.
In a model of 3-brane with matter  embedded in 5D space-time  
conditions for gravitons emitted from the brane to the bulk to return back
to the brane are found. 
For a given 5-momentum of graviton  falling back to the brane the 
interval between the times of emission and return to the brane is calculated.
A method to calculate contribution to the energy-momentum tensor from
multiple graviton bouncings  is developed.   
Explicit expressions for contributions to the  energy-momentum tensor
of gravitons which 
have made one, two and three bounces are obtained and their magnitudes 
are numerically calculated. These expressions are used to solve the
evolution equation for dark radiation. A relation connecting
reheating temperature and the scale of extra dimension is obtained.
For the reheating temperature $T_R\sim 10^6 GeV$ we estimate the scale of extra
dimension $\m$ to be of order $10^{-9} GeV\,\,\, (\m^{-1}\sim 10^{-5} cm )$.

\end{abstract}

\section{Introduction}

Brane-world scenarios with the observable Universe located on a 3-brane embedded in a 
higher-dimensional space-time have attracted considerable interest recently. Such
models with matter on the brane can reproduce the main cosmological data \cite{maart,dab,hung2,mi1}. 

A general property of extra-dimensional models is that although ordinary matter is supposed 
to be confined 
to a brane, gravity propagates in the whole space-time. This entails the effect that gravitons 
produced in reactions of particles on the brane can escape to the bulk.  
  Graviton production is strong in the early hot Universe, 
and can alter the  time evolution of matter on the brane and,
in particular, the  primordial nucleosynthesis.

In this paper we calculate graviton production in a model of five-dimensional Universe 
with one large extra dimension. Matter is supposed to be confined to the 3D brane.
Time evolution of matter in this model is described by the generalized 
Friedmann equation $H^2 =\r^2 +2\m\r + \cdots$ \cite{binetr1,binetr2,maeda}. 
We consider the period of early cosmology, in which 
the term quadratic in energy density
is dominant, $\m/\r\ll 1$ ($\r$ is the normalized energy
density on the brane defined in (\ref{1.13}),\, $\m
=(-\Lambda/6)^{1/2}$, and $\Lambda$ is 5D cosmological
constant).

Because  the space-time is curved, a part of  gravitons emitted in the bulk can return back to 
the brane \cite{heb,lan1,lan2} and bounce again to the bulk.
In paper \cite{mi2} an analytical method to show that a bounce is possible was developed. 
In the present paper 
we investigate further conditions of a bounce. 
Solving the combined system of equations of trajectories of the brane  and  of emitted graviton,
we find conditions
 at which graviton can fall back  to the brane. 
We derive an equation for the interval of time between graviton emission 
and its return to the brane 
and develop a scheme to calculate times of returns of graviton to the brane 
for multiple bounces.
We show that in the period of early cosmology the ratio of times $t_0 /t_1$ of 
graviton emission $t_0$
and its return to the brane $t_1$
to a good approximation can be expressed as a function of $x=m (t_1 )/E(t_1)$, 
where $(E(t_1 ),\,m(t_1 ),\,{\bf p})$ are the components
of graviton 5-momentum at the time $t_1$ when graviton  returns to the brane.

As an application of the above results, using the distribution function of emitted gravitons of
paper \cite{lan2}, we calculate the the components of the energy-momentum tensor of bouncing gravitons
$T^{in, (k)}_{nn}$ ($n^A$ is transverse to the brane, k is a number of a bounce).
For the first three bounces we obtain explicit expressions for $T^{in, (k)}_{nn}$
and estimate their numerical magnitudes.
The expressions for the energy-momentum tensor of bouncing gravitons
 are used to solve the evolution equation of dark radiation 
\cite{lan2,mi3}.
Solving this equation, we find  a relation connecting 
 the reheating temperature of the Universe  $T_R$  and
the scale of the extra dimension $\m$. Qualitative constraints on $T_R$ and $\m$
are discussed. 

In Sect.2 we review two approaches to the 5D model.

In Sect.3 we solve geodesic equations for gravitons propagating in the bulk. 

In Sect.4 we solve the combined system of equations for
graviton and brane trajectories  and find conditions for return of graviton to the brane.
We calculate the interval of times between graviton emission and detection as a function of
graviton momentum at the time of detection.

In Sect. 5 consider multiple graviton bounces. We calculate the $(nn)$ components 
of the energy-momentum tensor of gravitons falling to the brane.

In Sect. 6 we present qualitative numerical analysis of the energy-momentum tensor and discuss  
solution of evolution equation for dark radiation.

\section{3-brane in 5D bulk }

We consider the 5D model with one 3D brane embedded in the
bulk. Matter is confined to the brane, gravity extends to the bulk.
In the leading approximation we neglect
graviton emission from the brane to the bulk.
The action is taken in the form
\be
\l{1.1}
S_5  =\f{1}{2\k^2}\left[\int\limits_\Sigma \,d^5 x\sqrt{-g^{(5)}}
(R^{(5)} -
2\L) + 2\int\limits_{\pa {\Sigma}}K \right] -\int\limits_{\pa
{\Sigma}}
\,d^4 x\sqrt{-g^{(4)}} \,\h{\s} -
\int\limits_{\pa
{\Sigma}}d^4 x \sqrt{-g^{(4)}}L_m ,
\ee
where $x_4 \equiv y$ is coordinate of the infinite extra dimension, $\k^2 =8\pi/M^3$ .

The 5D model can be treated in two alternative
approaches. 
In the first approach  metric is non-static, and the brane is located at a fixed position in
the extra dimension \cite{binetr1,binetr2}.
We consider the class of metrics of the form
\be
\l{1.2}
ds^2_5 =g^{(5)}_{AB}dx^A dx^B =
-n^2 (y,t)dt^2 + a^2 (y,t)\eta_{ab} dx^a dx^b +dy^2 
\ee
The brane is spatially flat and  located at $y=0$.
The freedom of parametrization of $t$, allows to set $n(0,t)=0$.
The energy-momentum tensor of matter on the brane  is taken
in the form
\be
\l{1.4a}
\h{t}_\m^\n=diag\,\d (y)\{-\h{\r},\h{p},\h{p},\h{p}\}.
\ee
For the following it is convenient to introduce
the normalized expressions for energy
density, pressure and
cosmological constant on the brane which all have the same
dimensionality $[GeV]$
\be
\l{1.13}
\m =\sqrt{-\f{\L}{6}},\quad \s =\frac{\k^2\h{\s}}{6},
\quad \r =\frac{\k^2\h\r}{6}, \quad {p} =\frac{\k^2\h{p}}{6}.
\ee
Reduction of the metric (\ref{1.2}) to the brane is
\be
\l{1.3}
ds^2 =dt^2 + a^2 (0,t)\eta_{ab} dx^a dx^b .
\ee

The function $a (t)=  a(0,t)$ satisfies the generalized Friedmann
equation \cite{binetr2}
\be
\l{1.7}
H^2 (t) = -\m^2 +(\r +\s )^2 +\m \r_w (t)
,\ee
where $ H(t) =\dot{a}(t)/{a (t)}$ and $\r_w(t)$ is the  Weyl 
radiation term  \cite{maart}, which below is set to zero.

In the second approach the brane separates
two static 5D AdS spaces attached to both sides of the brane.
The metrics of the AdS spaces are solutions of the Einstein equations of the form
\be
\l{1.6}
ds^2 = -f_i (R) d\t{T}^2 +\f{d R^2}{f_i (R)} +\m_i^2 R^2 dx^a dx_a
,\ee
where
$$
f_i (R)= \m_i^2 R^2 - \f{P_i}{R^2}
.$$
Below we consider the case $\m_1 =\m_2$ and $P_i =0 $.
Trajectory of the moving brane in the $R,T$ plane is given by parametric
equations $R= r_b (t),\,\,\t{T}= \tau_b (t)$, 
where $t$ is the proper time on the brane
\be
\l{1.10}
 -f(r_b )\dot{\tau}_b^2 +f^{-1}(r_b )\dot{r}_b^2 =-1 ,
\ee
dot is derivative over $t$.
Reduction of the 5D metric to the brane is 
\be
\l{1.4}
ds^2 =-dt^2 +r_b^2 (t) dx^a dx_a.
\ee
The function $r_b (t)$ satisfies the generalized Friedmann equation \cite{kraus,col,cham}
\be
\l{1.8}
\left(\f{\dot{r_b}}{r_b}\right)^2 =-\m^2 +(\r +\s )^2
.\ee
Equations (\ref{1.8}) and (\ref{1.7}) with $\r_w (t)=0$ are of the same form, and
 $ a^2(0,t)$ can be identified with $  r_b^2 (t)$.
Below we consider the case $\s =\m$ \cite{mi1}, so that (\ref{1.8}) takes a form
\be
\l{1.9}
\left(\f{\dot{r}_b}{r_b}\right)^2 =\r^2 +2\m\r .
\ee
The normalized velocity vector of the brane and
the normal vector to the brane are
\be
\l{1.11}
v^A =(v^T ,\,v^R )=(\dot{\tau}_b ,\,\dot{r}_b),\qquad
n^A = (n^T ,\,n^R )=\pm \left(\f{\dot{r}_b}{f(r_b )} ,\,f(r_b )\dot{\tau}_b \right)
\ee
Here
\be
\l{1.12}
\dot{\tau}_b =\e\f{\sqrt{f+\dot{r}_b^2}}{f(r_b )},
\ee
where $\e=\pm$.

In the following we choose $n^A$ with the sign (-)  
$$
n^A = -\left(\f{\dot{r}_b}{f(a )} ,\,f(a )\dot{\tau}_b \right).
$$

\section{Geodesic equations in the picture with static metric}

Let $\la$ be  parameter along a geodesic. Geodesic equations in the metric (\ref{1.6}) are
\ba
\l{a.6}
&{}&\f{d^2 \t{T}}{d\la^2}+2\G^T_{TR}\f{d\t{T}}{d\la}\f{dR}{d\la}=0
\\\l{a.7}
&{}&\f{d^2 x^a}{d\la^2}+ 2\G^a_{bR}\f{dx^b}{d\la}\f{dR}{d\la}=0
\\\l{a.8}
&{}&\f{d^2 R}{d\la^2}+ \G^R_{RR}\left(\f{dR}{d\la}\right)^2 +
\G^R_{TT}\left(\f{d\t{T}}{d\la}\right)^2
+\G^R_{ab}\f{dx^a}{d\la}\f{dx^b}{d\la}=0,
\ea
We consider solutions of the geodesic equations even in $\la$.
Integrating the geodesic equations, one  obtains \cite{mi3}
\be
\l{a.9}
\f{d\t{T}}{d\la}=\f{C^T}{\m^2 R^2},\qquad \f{dx^a}{d\la}=\f{C^a}{\m^2 R^2},
\qquad
\left(\f{dR}{d\la}\right)^2 =\m^2 R^2\, ({C^{R}})^2 +{C^{T}}^2 -{C^a}^2 
,\ee
where $ (C^T ,\,C^{a} , \,C^{R} ) $ are integration parameters.

Tangent vectors to a null geodesic satisfy the relation
\be
\l{a.14}
g_{AB}\f{dx^A}{d\la}\f{dx^B}{d\la}=0
,\ee
from which it follows  that  $C^R =0$.
Eqs. (\ref{a.6})-(\ref{a.9}) were solved with the initial condition
that $\t{T}(0)$ and $R(0)$ are located on the brane world sheet:
$\t{T}(0)= \tau_b (t_0 ), \, 
R(0)=r_b (t_0 )$. 
Here $t_0$ is the proper time of the point on the brane world sheet $\tau_b (t_0 ),\,r_b (t_0 )$ 
 at which the   geodesic begins, i.e. the time of the graviton emission.

The components of momentum of a graviton propagating along a null geodesic are proportional to the  
tangent vector to a null geodesic 
\be
\l{3.10}
({p}^{T} ,\,{p}^{R} ,\,{p}^{{a}})\sim\left(\f{C^T}{\m^2 R^2},\,\e_R ({C^T}^2 -{C^a}^2)^{1/2} 
,\,\f{C^a}{\m^2 R^2}  \right)
,\ee
where $\e_R =\pm$. Also we define $\e_T$ as $C^T =\e_T |C^T|$.
Expanding  the graviton momentum $p^A$ in the basis $(v^A (t), n^A (t),
e^A_{\b{a}}(t))$, where $e^A_{\b{a}} =\d^A_{\b{a}} /\m a$, we have
\be
\l{3.1}
p^A =Ev^A +mn^A +p^{\b{a}} e^A_{\b{a}} ,
\ee
where
\be
\l{3.5}
 p^A p_A =-E^2 +m^2 +{p^{\b{a}}}^2 =0
.\ee
The components of $p^A$ in the  two bases are connected as
\ba
\l{3.2a}
p^T =\f{E\e\sqrt{\m^2 +H^2} -m\,H}{\m^2 r_b},\quad
p^R = r_b (E H -m \e\sqrt{\m^2 +H^2}), \quad p^a=\f{p^{\b{a}}}{\m r_b}\\ \l{3.2b}
E=p^T\e\sqrt{\m^2 +H^2}r_b-\f{p^R H}{\m^2 r_b},\quad
m=p^T Hr_b - p^R\f{\e\sqrt{\m^2 +H^2}}{\m^2 r_b}.
\ea
The components $E$ and $m$ depend on $t$  through  $r_b (t)$.   

Introducing 
\be
\l{a.17}
\g =\f{{C^a}^2}{{C^T}^2}=1-\f{{C^R}^2}{{C^T}^2}
,\ee
and expressing $\g$ through $E$ and $m$, we have
\be
\l{3.3}
\g = 1-\f{1}{(\m^2 R^2 )^2}\f{{p^R}^2}{{p^T}^2}=1-\left(\f{EH-m\e\sqrt{\m^2+H^2}} 
{E\e\sqrt{\m^2+H^2}-mH}\right)^2\f{r_b^4 (t)}{R^4}.
\ee
If at a time $t$ graviton is on the brane world sheet $R=r_b (t)$ 
($t$ is a time of emission, $t_0$, or a time 
of  return of the graviton to the brane, $t_1$, we  
take the basis $(v^A, n^A,  e^A )(t)$ at the time $t$ and obtain $\g$ in a form  
\be
\l{3.3a}
\g = \f{\m^2 (E^2 -m^2 )}{(E\e\sqrt{\m^2+H^2}-m H)^2}
.\ee

\section{Bounce of massless particles in the period of early cosmology}

We consider the radiation-dominated period of the
early cosmology, when $\r/\m \gg 1$ or,equivalently, $\m t\ll 1$. 
Supposing that the energy loss from the brane to the bulk
is sufficiently small to comply with the observational data, we neglect in the
conservation equation for the energy-momentum tensor the energy flow in the bulk.
In the period of early cosmology, in the model with extra dimension, from
the expression for energy density of relativistic degrees of freedom
\be
\l{T1}
\r (T) =\f{\k^2\pi^2 g_* (T)\,T^4}{180},
\ee
($g_* (T)$ is a total number of relativistic degrees of freedom \cite{kolb, rub} )
it follows that
\be
\l{T3}
\f{\r (t)}{\r (t_1 )}=\f{g_* (T)T^4}{g_* (T_1 )T^4_1} \simeq\f{T^4}{T_1^4}
\ee
For times $t_1$ and $t$ in the region of early cosmology, from the Friedman equation one obtains
\be
\l{T4}
\f{\r (t)}{\r (t_1 )}=\left(\f{r_b (t_1 )}{r_b (t )}\right)^4\simeq\f{t_1}{t}.
\ee

Following \cite{lan2}, with the use of (\ref{1.10}), the equation for the brane trajectory can be 
written as
\be
\l{2.2}
\f{dr_b}{d\tau_b}=\e\f{\m^2 r_b^2 \dot{r_b}}{\sqrt{\m^2 r_b^2 +\dot{r_b}^2}}=\e\m^2
r_b^2 \f{H}{\sqrt{\m^2 +H^2}}
.\ee
Integrating Eq. (\ref{2.2})  with the boundary
conditions $r_b =r_b (t_0 ),\,\,\, 
\tau_b =\tau_b (t_0)$, we obtain the equation for trajectory of the brane 
\be
\l{2.21}
\m^2 (\tau_b (t) -\tau_{b}(t_0))=\e\int_{r_b (t_0 )}^{r_b (t)} \f{dr} {r^2} \f{\sqrt{\m^2 +H^2}}{H}=
\e\left(\f{1}{r_b (t_0 )}-\f{1}{r_b (t )}\right)+
\e\int_{r_b (t_0 )}^{r_b (t)} \f{dr}{ r^2}\left[ \f{\sqrt{\m^2 +H^2}}{H}-1\right].
\ee
From the first integrals of the null geodesic equations (\ref{a.9})
we obtain
\be
\l{2.5}
\f{d{R}}{d{\t{T}}} =\e_T\e_R (1-\g )^{1/2}\m^2 {R}^2
.\ee
Integrating Eq. (\ref{2.5}) with the initial conditions $R =r_b (t_0),\, \t{T} =\tau_b
(t_0)$,
we obtain the equation for for a null geodesic (graviton trajectory)
\be
\l{2.6}
 \f{1}{r_b(t_0)}-\f{1}{R} =\e_T\e_R  (1-\g )^{1/2}\m^2 (\t{T} -\tau_b (t_0) )
.\ee
If graviton returns to the brane at time $t_1$, we have $R=r_b (t_1 )$.
Combining Eqs. (\ref{2.21}) and (\ref{2.6}) and using 
 Friedmann equation,  $H^2 =\r^2 +2\m\r$, we obtain an equation for $r_b (t_1 )$
\be  
\l{2.3}
[\e \e_T \e_R (1-\g )^{-1/2} -1]\left(\f{1}{r_b (t_0 )}-\f{1}{r_b (t_1 )}\right)=
\int_{r_b (t_0 )}^{r_b (t_1 )} \f{dr} {r^2} \left[\f{\r +\m}{\sqrt{\r^2 +2\m\r }}-1\right]
.\ee
Eq. (\ref{2.3}) can be interpreted as an equation which determins the time of return of graviton
to the brane $t_1$ for a given time of emission $t_0$. 
It is seen that Eq.(\ref{2.3}) admits solution only if
\be
\l{2.61}
\e\e_T\e_R =\,( +)
\ee
 Expanding the integrand of 
(\ref{2.3}) in powers of $\m/\r$, we have
$$
\f{\r+\m}{\sqrt{\r^2+2\r\m}}=
 \left(1+\f{1}{2}\left(\f{\m}{\r}\right)^2 - 
\left(\f{\m}{\r}\right)^3 + \f{15}{8}\left(\f{\m}{\r}\right)^4 +...\right),
$$
where  $\r (t) =\r (t_0)(r_b (t_0)/r_b (t))^4$.  
Integrating Eq. (\ref{2.3}),  we obtain
\ba
\l{2.4}
&{}&((1-\g )^{-1/2} -1)\left(\f{1}{r_b (t_0 )}-\f{1}{r_b (t_1 )}\right)\\\nonumber
&{}&= \f{1}{r_b(t_0 )}
\left[\f{1}{14}\left(\f{\m}{\r_0}\right)^2\left( \left(\f{r_b (t_1  )}{r_b (t_0 )}\right)^7-1\right)
-\f{1}{11}\left(\f{\m}{\r_0}\right)^3
\left(\left(\f{r_b (t_1 )}{r_b (t_0 )}\right)^{11}-1\right)
+ ...\right]
.\ea
The series in (\ref{2.4}) is convergent.
Introducing
\be
\l{2.71}
z=\f{r_b (t_0)}{r_b (t_1 )}\simeq \left(\f{t_0}{t_1}\right)^{1/4}
\ee 
and substituting $\r_0 =\r_1 z^{-4}$, where $ \r_0=\r (t_0 )$ and $\r_1=\r (t_1 )$, 
we transform (\ref{2.4}) to a form
\be
\l{2.7}
[ (1-\g )^{-1/2}-1]=
\f{1}{2}\left(\f{\m}{\r_1}\right)^2 \f{z(1-z^7 )}{7(1-z)}
-\left(\f{\m}{\r_1}\right)^3
\f{z(1- z^{11} )}{11(1-z)}+...
.\ee
The function
\be
\l{2.9}
f_k (z)=\f{z(1-z^k )}{k(1-z)}
\ee
is monotone increasing with the maximum at the point $z=1$ equal to $1$.

\subsection{Conditions of the fall of graviton on the brane}

The sign of the graviton momentum component $m(t)$ at the time  $t_1$ at which graviton returns
 to the brane is opposite to that at the time of emission $t_0$.
From (\ref{3.2b}) we have
\ba
\l{n1}
E\sim \e_T\sqrt{H^2+\m^2}-\e_R H (1-\g)^{1/2}\\\nonumber
m\sim \e_T H  - \e_R\sqrt{H^2+\m^2} (1-\g)^{1/2}
.\ea
Condition (\ref{2.61}) is satisfied in the following cases:
\ba
\nonumber
\mbox{(i)}\,\, \e_T =+,\,\,\,\e_R =+;\,\,\, \e = + ;
\qquad \mbox{(ii)} \e_T =-,\,\,\,\e_R =-;\,\,\, \e = +  ;\\\nonumber
\mbox{(iii)} \e_T =- ,\,\, \e_R =+ ;\,\,\, \e =-  ;
\qquad \mbox{(iv)} \e_T =+ ,\,\,\e_R =-	;\,\, \e =- 
\ea
Let us find in which case is realized one of the possibilities:
either $(a)\,\,\, m_0 = m(\tau_0)>0,\,\,\,m_1 =  m(t_1 )<0$, or $\quad  (b)\,\, m_0 <0,\,\,\, m_1 >0$.
\begin{itemize}
\item
The case (i)(a). $\e_T =\e_R =\e=+$ .

$ E $ and $m$ are 
$$
E(t)\sim \sqrt{H^2+\m^2}-H (1-\g)^{1/2} >0,\qquad
m(t)\sim H  - \sqrt{H^2+\m^2} (1-\g)^{1/2}.
$$
Conditions 
$$
m(\tau_0) \sim H_0  - \sqrt{H_0^2+\m^2} (1-\g)^{1/2}>0,\qquad
m(t_1 )\sim H_1  - \sqrt{H_1^2+\m^2} (1-\g)^{1/2}<0.
$$
 are satisfied, if
\be
\l{n2}
\f{H_0}{\sqrt{\m^2 +H_0^2}}>(1-\g)^{1/2}>\f{H_1}{\sqrt{\m^2 +H_1^2}}
.\ee
Here $H_0 =H(t_0),\,\,H_1 =H(t_1)$. 
\item The case (i)(b). $\e_T =\e_R =\e=+$.

$E$ and $m$ are
$$
E(t)\sim \sqrt{H^2+\m^2}-H (1-\g)^{1/ 2} >0 \qquad
m(t)\sim H  - \sqrt{H^2+\m^2} (1-\g)^{1/2}.
$$
 From conditions $m_0 <0,\,\,\, m_1 >0$ it follows that
$$
\f{H_0}{\sqrt{\m^2 +H_0^2}}<(1-\g)^{1/2}<\f{H_1}{\sqrt{\m^2 +H_1^2}}
$$
or $H_0<H_1$, which is impossible, because $t_0< t_1$.
\item The case (ii)(a). $\e_T =\e_R=- ,\,\,\,\e=+$ .

Analogously to the case (i)(b) in the case (ii)(a) there are no solutions. 
\item  The case (ii)(b). 	$\e_T =\e_R=- ,\,\,\,\e=+$  .

$ E $ and $m$ are 
$$
E(t)\sim -\sqrt{H^2+\m^2}+H (1-\g)^{1/2} <0\qquad
m(t)\sim -H  + \sqrt{H^2+\m^2} (1-\g)^{1/2}.
$$
Solution  $m_0 <0, \,\, m_1 >0 $ exists, provided (\ref{n2}) is valid.
\item The case (iii)(a).  $ \e_T =-,\,\,\,\e_R=+$.
 
 $ E$ and $m$ are
$$
E(t)\sim -\sqrt{H^2+\m^2}-H (1-\g)^{1/2} \qquad
m(t)\sim -H  - \sqrt{H^2+\m^2} (1-\g)^{1/2}<0. 
$$
Because $m$ is negative and  does not change its sign, there are no solutions.
Analogously in the case (iv) $m$ is always positive, and no solution exists. 
\end{itemize}

To conclude, we are left with the solutions of the types (i)(a) and (ii)(b), which are 
physically equivalent, because Eqs. (\ref{a.9}) with  $\e_T=\e_R=+$ transform 
to equations with $\e_T=\e_R=-$ under the change $\lambda \rightarrow -\lambda$.
In the following we consider the case (i)(a). 

\subsection{Relation between emission and detection times}

To solve Eq.(\ref{2.7}) we  need to transform the expression $(1-\g )^{-1/2} -1$
to a convenient form.
Introducing $x=E/m$ and using Friedman equation, $H^2 =\r^2 +2\m\r$, and (\ref{3.3a}),  we have
\be
\l{n3}
(1-\g )^{-1/2} -1=\left|\f{\sqrt{H^2+\m^2}-xH}{H -x\sqrt{H^2+\m^2}}\right|-1=
\left|\f{1+\m/\r-x\sqrt{1+2\m/\r}}{\sqrt{1+2\m/\r}-x(1+\m/\r)}\right|-1
\ee
Relation (\ref{n3}) is valid at the endpoints of graviton trajectory,
at emission point and at points where graviton hits the brane.

Because, as discussed in preceding subsection, $ x_1 <0$, 
the relation (\ref{n3}) written at  time $t_1$  takes the form 
\be
\l{3.11}
(1-\g )^{-1/2} -1=(1-|x_1|)\f{1+\m/\r_1 -\sqrt{1+2\m/\r_1}}{|x_1|(1+\m/\r_1 )+\sqrt{1+2\m/\r_1}}
.\ee
In the period of early cosmology expression (\ref{3.11}) approximately is
\be
\l{3.12}
(1-\g )^{-1/2} -1\simeq(1-|x_1|)\f{\m^2/2\r_1^2}{\sqrt{1+2\m/\r_1}+  |x_1| (1+\m/\r_1 )}.
\ee
The advantage of this form of $\g$ is that we have extracted the factor $(\m/\r_1 )^2$.
 Now the Eq. (\ref{2.7}) can be  written as
\be
\l{2.7a}
\left(\f{\m}{\r_1 }\right)^2\f{1-|x_1|}{(1+|x_1|)(1+\m/\r_1 )}=\left(\f{\m}{\r_1 }\right)^2
\left[f_7 (z_{01})-\left(\f{\m}{\r_1}\right)2f_{11}(z_{01}) +...\right],
\ee
or
\be
\l{2.7b}
\f{1-|x_1|}{(1+|x_1|)}=
\left[f_7 (z_{01})-\left(\f{\m}{\r_1}\right)2f_{11}(z_{01}) +...\right](1+\m/\r_1 ),
\ee
Eqs. (\ref{2.7}) define $z_{01}\simeq (t_0/t_1)^{1/4}$ through the ratio $x_1=m_1 /E_1$ 
and $\r_1 =\r (t_1 )$, or,
equivalently, the  emission time $t_0$ through the "mass" and "energy" of the graviton
at the time of return of the graviton to the brane $t_1$.
In the following, for practical calculations, in the region $\m/\r_1< 1$
we use a simplified equation
\be
\l{2.8}
\f{1-|x_1|}{1+|x_1|}\simeq f_7 (z_{01}).
\ee
Domain of applicability and corrections to this equation are discussed in Sect.6.


\section{Multiple bounces}

To consider  multiple reflections from the brane of bouncing gravitons we use  
matrix notations. 
Introducing
\ba
\l{4.1}
K=K^{-1}=\left| \begin{array}{cc}
\sqrt{H^2/\m^2+1} &-H/\m\\[2mm]H/\m&-\sqrt{H^2/\m^2 +1}
\end{array}
\right|
,\ea
and
\be
\l{4.01}
\t{p}^T= p^{T}\m R, \qquad \t{p}^{R}=\f{p^R}{\m R},\quad {p^{\b{a}}}=\m R p^a
,\ee
we have
\ba
\l{4.2}
\left(
\begin{array}{c}
  \t{p}^{T}\\ \t{p}^{R}
\end{array}
\right)=
K\left(
\begin{array}{c}
 E \\ m
\end{array}
\right),
\qquad
\left(
\begin{array}{c}
  E\\ m
\end{array}
\right)=
K
\left(
\begin{array}{c}
  \t{p}^{T}\\ \t{p}^{R}
\end{array}
\right).
\ea

The case with multiple bouncings is illustrated by the scheme
\ba
\l{4.4}
\left(\begin{array}{c}p_0\\E_0\\m_0\end{array}\right)^{out}(t_0)
\stackrel{z_{01}}{\rightarrow}
\left(\begin{array}{c}p_1^{in}=
p_1^{out}\\
E_1^{in}=E_1^{out}\\
m_1^{in}=-m_1^{out}\end{array}\right)(t_1 )
\stackrel{z_{12}}{\rightarrow}
\left(\begin{array}{c} p_2^{in}=
p_2^{out}\\
E_2^{in}=E_2^{out}\\
m_2^{in}=-m_2^{out}\end{array}\right)(t_2 )
\stackrel{z_{23}}{\rightarrow} 
\cdots
\ea
The left column corresponds to the emission time $t_0$. 
 In the next brackets, in the left columns are momenta of 
 ingoing particle, in the right
ones are the outgoing.
Under reflection from the brane momentum $p^a$ parallel to the brane and energy $E$ are
conserved, 
transverse momentum to the brane $m$ changes its sign:
$$
E\rightarrow E, \qquad p^a\rightarrow p^a, \qquad m\rightarrow -m.
$$
Momenta $\t{p}^{T,R}\sim C^{T,R}/r_b (t )$
are are rescaled when moving from one bracket to the next
\be
\l{4.81}
\f{\t{p}^{T,R}_n}{\t{p}^{T,R}_{n-1}}=\f{r_b (t_{n-1} )}{r_b (t_n )} 
\simeq\left(\f{t_{n-1}}{t_n}\right)^{1/4}= z_{n-1,n}.
\ee
Transformation from "in" to "out"  components within a bracket is given by
\be
\l{4.5}
M^{out}\equiv\left(
\begin{array}{c}
 E^{out} \\m^{out}
\end{array}
\right)
=\left(
\begin{array}{cc}1&0\\0&-1\end{array}\right)
\left(
\begin{array}{c}
 E^{in} \\m^{in}
\end{array}\right)\equiv L M^{in}
.\ee
Let us consider the first two brackets in  (\ref{4.4}).  
Using (\ref{4.2}) we express $E_0$ and $m_0$ through  $E_1$ and $m_1$
and obtain
\ba
\l{4.9}
&{}&\left(
\begin{array}{c}
 E_0 \\ m_0
\end{array}\right)^{out}= z_{01}^{-1}
K_0 K_1
\left(
\begin{array}{c}
 E_1 \\ m_1
\end{array}\right)^{in}
\\\nonumber
&{}&= z_{01}^{-1}
\left(
\begin{array}{cc} {\sqrt{\f{H_0^2}{\m^2} +1}\sqrt{\f{H_1^2}{\m^2} +1}-\f{H_0 H_1}{\m^2}} &
{\f{H_0}{\m}\sqrt{\f{H_1^2}{\m^2} +1}- \f{H_1}{\m}\sqrt{\f{H_0^2}{m^2} +1}}
 \\[3mm]{\f{H_0}{\m}\sqrt{\f{H_1^2}{\m^2} +1}- \f{H_1}{\m}\sqrt{\f{H_0^2}{\m^2} +1}} &
{\sqrt{\f{H_0^2}{\m^2} +1}\sqrt{\f{H_1^2}{\m^2} +1}-\f{H_0 H_1}{\m^2}}
\end{array}\right)
\left(
\begin{array}{c}
 E_1 \\[3mm] m_1
\end{array}\right)
.\ea
In the  period of early cosmology,
  from the Friedman equation it follows that $H/\m \simeq 1/(4\m t)$.
For small $\m t$ we can simplify the expressions in (\ref{4.9}) as
\ba
\l{4.14}
{\sqrt{\f{H_0^2}{\m^2} +1}\sqrt{\f{H_1^2}{\m^2} +1}-\f{H_0 H_1}{\m^2}}
\simeq
\f{1}{2}\left(\f{H_0}{H_1}+\f{H_1}{H_0}\right)
\simeq \f{1}{2}\left(\f{t_1}{t_0} +\f{t_0}{t_1}\right)=\f{z_{01}^{-4}+z_{01}^{4}}{2}\\\nonumber
{\f{H_0}{\m}\sqrt{\f{H_1^2}{\m^2} +1}- \f{H_1}{\m}\sqrt{\f{H_0^2}{\m^2} +1}}
\simeq
\f{1}{2}\left(\f{H_0}{H_1}-\f{H_1}{H_0}\right)
\simeq\f{1}{2} \left(\f{t_1}{t_0} - \f{t_0}{t_1}\right)=\f{z_{01}^{-4}-z_{01}^{4}}{2}
.\ea
Introducing
$$
\psi^{\pm}(z)=\f{z^{-4}\pm z^4}{2}
,$$
 we obtain
\be
\l{4.15}
 \left(\begin{array}{c} E_0 \\ m_0\end{array}\right)^{out}= z_{01}^{-1}
\left(\begin{array}{cc}\psi^+_{01}&\psi^-_{01}\\
\psi^-_{01}&\psi^+_{01}\end{array}\right)
\left(\begin{array}{c} E_1 \\ m_1\end{array}\right)^{in}
,\ee
where
$$
\psi^\pm_{01}=\psi^\pm (z_{01} )
$$
For multiple bouncings we have
\ba
\l{4.6}
&{}&M_0^{out}=z_{01}^{-1}K_0 K_1 M_1^{in},\\\nonumber
&{}&M_0^{out}=z_{02}^{-1}K_0\,K_1 LK_1\, K_2 M_2^{in}, \\\nonumber
&{}&\cdots\\\nonumber
&{}&M_0^{out}={Z}_{0,n}^{-1}K_0\,K_1 LK_1\, ...K_{n-1}K_n M^{in}_n
,\ea
where
\ba
\l{4.7}
&{}&z_{0n}=z_{01}\,z_{12}\,z_{23}\cdots z_{n-1,n},\\
[2mm]
\l{4.7a}
&{}&K_0\,K_1 LK_1\, ...K_{n-1}K_n=
\left(\begin{array}{cc}\psi^+_{0n}&\psi^-_{0n}\\
[2mm]
(-)^{n+1}\psi^-_{0n}&(-)^{n+1}\psi^+_{0n}\end{array}\right).
\ea
Here
\ba
\l{4.8a}
&{}&\psi^\pm_{0n}=\psi^\pm (u_{0n} ),\\
[2mm]
\l{4.8b}
&{}&u_{0n}=u_{0,n-1}^{-1}\,z_{n-1,n}
,\ea
or explicitly
$$
u_{02}=z_{01}^{-1}\,z_{12}, \quad u_{03}=z_{01}\,z_{12}^{-1}\,z_{23},\quad 
u_{04}=z_{01}^{-1}\,z_{12}\,z_{23}^{-1}z_{34},\,\cdots
$$
From (\ref{4.8a}) and (\ref{4.8b}) it follows that $u_{0,2k+1}<1$ and  $u_{0,2k}>1$ . 
In the latter case
\be
\l{5.20}
\psi^-_{0,2k}(u_{0,2k})=-|\psi^-_{0,2k}(u_{0,2k})|=-\psi^-_{0,2k}(u_{0,2k}^{-1})
,\ee
and
\ba
\nonumber
E_0 =z_{0,2k} (\psi^+_{0,2k}E_{2k} - |\psi^-_{0,2k}|m_{2k})\\
[2mm]\nonumber
m_0=z_{0,2k} (|\psi^-_{0,2k}|E_{2k} - \psi^+_{0,2k}m_{2k}) .
\ea
It should be noted that the functions $z_{n-1,n}$ in processes with different number of
bounces are different. If $z^{(k)}_{n-1,n}$ refers to the process with
$k-1$ bounces, different $z^{(k)}_{n-1,n}$  are connected by the following  relations
\ba
\l{5.19}
&{}&z^{(k)}_{k-1,k}=z^{(k-1)}_{k-2,k-1}=z^{(k-2)}_{k-3,k-2}=\cdots\,= z^{(1)}_{01},\\\nonumber
&{}&z^{(k)}_{k-2,k-1}=z^{(k-1)}_{k-3,k-2}=z^{(k-2)}_{k-4,k-3}=\cdots \,= z^{(2)}_{01},\\\nonumber
&{}&\cdots \qquad\cdots\\\nonumber
&{}&z^{(k)}_{12}=z^{(k-1)}_{01}
.\ea

The distribution function of non-interacting gravitons  
in the bulk satisfies the Liouville equation
 without the collision term. If coordinates and momenta gravitons along a geodesic are parametrized
by parameter $\la$, i.e. $f(x^A (\la ), p^A (\la ))$, we have
  \be
\l{5.1a}
f (R (\la_0 ), {p}^A (\la_0 ))= f (R (\la_1 ), {p}^A (\la_1 ) )
.\ee
In the case of the metric (\ref{1.6}) relation (\ref{5.1a}) can be written as \cite{lan2}
$$
f(\t{T}_0, R_0 , \t{p}^A )=f(\t{T} , R_1,  \t{p}^A R_1/R_0 )
,$$   
where $\t{p}^A = (\t{p}^T ,\t{p}^R ,  {\bf p})$. If the points 
$(\t{T}_0, R_0 )$ and $(\t{T}_1 , R_1 )$
are on the brane world sheet, they are functions of the proper time on the 
brane. Temperature of the Universe, $T(t)$, 
is defined through the proper time $t$ via (\ref{T3})-(\ref{T4}).
   
\subsection{First fall of gravitons to the brane}

We suppose that the distribution function of emitted gravitons $f^{out}$ 
 depends on $E_0=\sqrt{m^2_0 + {\bf p}^2}, m_0 $ and temperature $T_0$, i.e.
$f^{out}=f^{out} (E_0, m_0, T(t_0 ))$.
 The distribution function of gravitons emitted at time $t_0$ and falling back
for the first time on the brane at time $t_1$ is
\be
\l{c1}
f^{in,(1)}(E_1 ,m_1 , T ) =
f^{out}(E_0 (E_1 ,m_1, z_{01} ), m_0 (E_1 ,m_1, z_{01} ), T_0 (T, z_{01} ))
,\ee    
where
\ba
\l{4.16}
E_0 =z^{-1}_{01}(\psi^+_{01} E_1 +\psi^-_{01} m_1)\\
[2mm]\l{4.16a}
m_0=z^{-1}_{01}(\psi^-_{01} E_1 +\psi^+_{01} m_1)
.\ea
 
In the period of early cosmology from (\ref{T3}) and (\ref{T4}) it follows that
$T_1/T_0\simeq \r_b (t_0)/\r_b (t_1)=z_{01}$. We obtain
the distribution function at time $t_1$ as  
\be
\l{c2}
f^{in,(1)} (E_1, m_1, T ) =f^{out} (z_{01}^{-1} (E_1\psi^+ (z_{01} )+m_1\psi^- (z_{01} )),
z^{-1}_{01}( E_1\psi^- (z_{01})+m_1 \psi^+ (z_{01})), z^{-1}_{01}T)
,\ee
where $z_{01}$ is determined as a function of $x_1=m_1/E_1 $ by Eq.(\ref{2.7a}).
For $\m/\r_1\ll 1$
 the terms $O(\m/\r_1 )$ could be neglected and $z_{01}$ is defined via (\ref{2.8}). 

Condition that $x_0=m_0/E_0>0$ takes the form
$$
x_0 =\f{\psi^-(z_{01})-|x_1|\psi^+(z_{01}) }{\psi^+(z_{01}) -|x_1| \psi^-(z_{01})}>0
.$$
Because the nominator of this ratio is positive, this condition is equivalent to
$\psi^-(z_{01})-|x_1|\psi^+(z_{01})>0$, or
$$
\f{1+|x_1|}{1-|x_1 |}z_{01}^8<1.
$$ 
To show that this inequality is satisfied, we wright
\be
\l{4.17}
1=\f{1+|x_1|}{1-|x_1|}f_7 (z_{01})=
\f{1+|x_1|}{1-|x_1|}z_{01}\f{1+...+z_{01}^6}{7}>\f{1+|x_1|}{1-|x_1|}z_{01}^8.
\ee

\subsection{Next falls of gravitons to the brane}  

The distribution function of gravitons emitted at time $t_0$, which bounced off 
the brane at time $t_1$
and fall on the brane  the second time  at time $t_2$  is
\ba
\l{5.1b}
f^{in,(2)}(E_2, m_2, t_2 )=f^{out,(1)}(E_1 (E_2 , m_2 ), - m_1 (E_2 , m_2 ), t_1 )
=f^{in,(1)}(E_1 (E_2 , m_2 ), m_1 (E_2 , m_2 ), t_1 )
\\\nonumber = f^{out}(E_0 (E_1 (E_2 , m_2 ), m_1 (E_2 , m_2 )),\,
m_0 (E_1 (E_2 , m_2 ), m_1 (E_2 , m_2 )),\, t_0).
\ea
Tracing the propagation of graviton, we obtain 
\ba
\l{5.21}
&{}&\left(\begin{array}{c}E_0 \\ m_0
\end{array}\right)^{out}
=z_{01}^{-1}\left(\begin{array}{cc}\psi_{01}^+ &\psi_{01}^-\\ 
\psi_{01}^- &\psi_{01}^+\end{array}\right)
\left(\begin{array}{c} E_1 \\ m_1 \end{array}\right)^{in}\\\nonumber&{}&
=z_{01}^{-1}\left(\begin{array}{cc}\psi_{01}^+ &\psi_{01}^-\\
 \psi_{01}^- &\psi_{01}^+\end{array}\right)
z_{12}^{-1}\left(\begin{array}{cc}\psi_{12}^+ &\psi_{12}^-\\
 -\psi_{12}^- &-\psi_{12}^+\end{array}\right)
\left(\begin{array}{c}E_2 \\ m_2 \end{array}\right)^{in}
=(z_{01}z_{12})^{-1}\left(\begin{array}{cc}\psi_{02}^+ &\psi_{02}^-
\\ \psi_{02}^- &\psi_{02}^+\end{array}\right)
\left(\begin{array}{c} E_2 \\ m_2 \end{array}\right)^{in}
,\ea
where $u_{02}=z_{01}z_{12}^{-1}$ and
$$
\psi^\pm_{02}=\f{1}{2}\left[\psi^+(z_{01})\psi^\pm (z_{12})-\psi^-(z_{01})\psi^\mp (z_{12})\right]=
 \f{1}{2}\left[ u_{02}^{-4} \pm u_{02} ^{4} \right].
$$
The time  $t_1$ of the bounce is determined from Eq. (\ref{2.8}) with 
$z=z_{12}\simeq (t_1/t_2)^{1/4}$, 
\be
\l{5.7}
\f{1-|x_2| }{1+|x_2| }= f_7 (z_{12}),
\ee
where $x_2 =(m_2/E_2)^{in}$.
The time $t_0$ is determined from the equation (\ref{2.8}) with $z=z_{01}\simeq (t_0/t_1)^{1/4}$
\be
\l{5.8}
\f{1-|x_1| }{1+|x_1| }= f_7 (z_{01}),
\ee
where $x_1 =(m_1/E_1)^{in}$. Expressing $x_1$ through $x_2$, we have
\be
\l{5.9}
x_1=x_1^{in}=-\f{\psi^- (z_{12}) +x_2 \psi^+ (z_{12})}{\psi^+ (z_{12})+x_2 \psi^- (z_{12})}=
-\f{\psi^- (z_{12}) -|x_2| \psi^+ (z_{12})}{\psi^+ (z_{12})-|x_2| \psi^- (z_{12})}.
\ee
Substituting (\ref{5.9}) in (\ref{5.8}), we obtain
\be
\l{5.12}
f_7 (z_{01})=\f{1-|x_1|}{1+|x_1|} =
\f{\psi^+(z_{12})-|x_2| \psi^-(z_{12})-\psi^-(z_{12})+|x_2| \psi^+(z_{12})}
{\psi^+(z_{12})-|x_2| \psi^-(z_{12})+\psi^-(z_{12})-|x_2| \psi^+(z_{12})} =
\f{1+|x_2|}{1-|x_2|}z_{12}^{8}.
\ee
Condition $m_1^{out}>0$ yields the constraint $\psi^- (z_{12})-|x_2| \psi^+ (z_{12})>0$. 
This condition can be rewritten as  $(1-z_{12}^8 )/(1+z_{12}^8 )>|x_2|$, or equivalently, as
$$
z_{12}^8 \f{(1+|x_2|)}{(1-|x_2|)}<1
,$$ 
which is valid, because of (\ref{5.12}). 

Condition $m_0 >0$ is $-(\psi_{02}^{-}-|x_2|\psi_{02}^{+}) >0$, 
or $z_{01}^8 (1-|x_2|)<z_{12}^8 (1+|x_2|)$,
which is satisfied, because 
$$
1<\f{1+|x_2|}{1-|x_2|}\f{z_{12}^8}{z_{01}^8}=\f{f_7 (z_{01})}{z_{01}^8}.
$$ 
Using relations (\ref{5.7}) and (\ref{5.12}) we can show that $z_{12}>z_{01}$.

The distribution function of gravitons (\ref{5.1b}) cab be expressed as
\ba
\l{5.10}
f^{(2)}(E_2, m_2, T )
 =f^{out, (0)}(z_{02}^{-1}
(-E_2\psi^-_{02} -m_2\psi^+_{02}),\,z_{02}^{-1}(E_2\psi^+_{02} +m_2\psi^-_{02}),\, 
z_{02}^{-1}T )
.\ea

In the case that gravitons have made two bounces, using (\ref{4.6}) and (\ref{4.7}), we have
\be
\l{5.13}
\left(\begin{array}{c}
 E_0 \\ m_0
\end{array}\right)^{out}= z_{03}^{-1}K_0 K_1 L K_1 K_2 L K_2 K_3
\left(\begin{array}{c}
 E_3 \\ m_3
\end{array}\right)^{in}= z_{03}^{-1}
\left(\begin{array}{cc} \psi^+_{03} & \psi^-_{03} \\ \psi^-_{03}&\psi^+_{03}
\end{array}\right)
\left(\begin{array}{c}
 E_3 \\ m_3
\end{array}\right).
\ee
Here $z_{03} = z_{01}z_{12}z_{23},\,\, \psi^\pm_{03}=\psi^\pm (u_{03}) $ 
and $u_{03}=z_{01}z_{12}^{-1}z_{23}$.
The functions $z_{k-1,k}$ are determined from the equations
\ba
\l{5.15}
f_7 (z_{23})=\f{1-|x_3|}{1+|x_3|},\\
\l{5.16}
f_7 (z_{12})=\f{1+|x_3|}{1-|x_3|}z^8_{23},\\
\l{5.17}
f_7 (z_{01})=\f{1-|x_3|}{1+|x_3|}\f{z_{12}^8}{z^8_{23}}.
\ea
 Using the above relations, we can show that $z_{23}>z_{12}>z_{01}$.

The distribution function $f^{(3)}(m_3, E_3, T )$ is
\ba
\l{5.18}
f^{(3)}(m_3, E_3, T )
 =f^{out, (0)}(z_{03}^{-1}
(E_3\psi^-_{03} +m_3\psi^+_{03}),\,z_{03}^{-1}(E_3\psi^+_{03} +m_3\psi^-_{03}),\,
z_{03}^{-1}T )
\ea

\section{Numerical estimates and discussion}

We perform numerical estimates of the $(nn)$ component of the energy-momentum tensor of incoming
gravitons using the distribution function of paper \cite{lan2}.
Qualitatively, the energy-momentum tensor of incoming gravitons at the registration time $t_1$ 
is formed by summing contributions from gravitons emitted at times $t_0$ preceding the
registration time $t_1$.

The distribution function of emitted gravitons is 
\be
\l{4.91}
 f^{(0)}(m,{\bf p},t_0 )=B m^3 e^{-E/T_0},\qquad B=\f{A\k^2}{2^{10}\pi^5},
\ee
where $E=\sqrt{m^2 +{\bf p}^2},\,\,\,A$ is the weighted sum of relativistic degrees of freedom
which contribute to the annihilation amplitude to gravitons \cite{heb}.
  The $(nn)$ component of the energy-momentum tensor
of emitted gravitons,  $T^{(em)}_{nn} (t_0 )$, is
\be
\l{T}
T^{em}_{nn}(t_0 ) = \int dm d{\bf p}\f{m^2}{2E}f^{(0)}(m , {\bf p}, t_0 )
.\ee
The $(nn)$ component of the energy-momentum tensor of gravitons falling back to the brane is 
\be
\l{Ta}
T^{in,(1)}_{nn}(t_1 )=\int dm_1 d{\bf p_1}\f{m_1^2}{2E_1}f^{(1)}(m_1 , {\bf p_1}, t_1 ),
\ee
where the distribution function of infalling gravitons is
\be
\l{T2}
f^{in,(1)}(m_1 , {\bf p_1}, t_1, T )=Bm_0^3 (m_1, E_1, z_{01} )
\exp{\left\{-\f{E_0 (m_1, E_1, z_{01} )}{T_0 (T, z_{01} )}\right\}}
.\ee
 $E_0 (m_1, E_1, z_{01} )$ and $m_0 (m_1, E_1, z_{01} )$ are defined by (\ref{4.16})-(\ref{4.16a}).
Substituting these expressions, we have
\ba
\l{4.15a}
T_{nn}^{in,(1)}=
2\pi B \int_{-E_1}^0\,dm_1 \,m_1^2\,\sqrt{E_1^2 -m_1^2}\int dE_1
z_{01}^{-3}( E_1 \psi^-_{01} +m_1 \psi^+_{01} )^3
\exp{ \left\{- \f{E_1 \psi^+_{01} + m_1 \psi^-_{01} }{T} \right\}} \\\nonumber
=2\pi B\int_0^{E_1}\,dm_1 \,m_1^2\,\sqrt{E_1^2 -m_1^2}\int dE_1
z_{01}^{-3}( E_1 \psi^-_{01} -m_1 \psi^+_{01} )^3
\exp{ \left\{- \f{E_1 \psi^+_{01} - m_1 \psi^-_{01}}{T} \right\}}
\ea
Introducing  $x =m_1/E_1$, we express $T^{(1)}_{nn}$ as
\be
\l{5.1}
T_{nn}^{in,(1)}=2\pi B\int dx x^2 \sqrt{1-x^2}
\int dE_1 E_1^7 (\psi^-_{01} -x \psi^+_{01} )^3 z_{01}^{-3}(x)
\exp\left\{\f{-E_1 (\psi^+_{01} -x \psi^-_{01})}{T}\right\}.
\ee
First, we integrate over $E_1$ in the limits $(0,\infty )$, and below we consider 
integration taking into account lower and upper bounds. For $T^{(1)}_{nn}$ we have
\be
\l{5.2}
T_{nn}^{in,(1)}( T) =2\pi B T^8 \Gamma (8)\int_0^1 dx x^2 \sqrt{1-x^2}z^{-3}_{01}(x)
\f{(\psi^-_{01}-x \psi^+_{01}  )^3 }
{(\psi^+_{01}  - x\psi^-_{01})^8}  
\ee
For the energy-momentum tensor of gravitons which have made 
one bounce we obtain
\be
\l{5.3}
T_{nn}^{in,(2)}
=2\pi^2 B\int^{\infty}_0\,dE_2 \int_{-E_2}^0 dm_2 \,m_2^2\,\sqrt{E_2^2 -m_2^2}\,
z_{02}^{-3}( -E_2 \psi^-_{02} -m_2 \psi^+_{02} )^3
\exp{ \left\{- \f{E_2 \psi^+_{02} + m_2 \psi^-_{02} }{T} \right\}} 
,\ee
where $z_{02}=z^{(2)}_{12}\,z^{(2)}_{01}$.
From the inequality $z_{12}>z_{01}$ it follows that  $\psi_{02} (z^{(2)}_{12}/z^{(2)}_{01})<0$.
Instead, we use $|\psi_{02}| = \psi_{02} (z^{(2)}_{01}/z^{(2)}_{12})$.
$T_{nn}^{in,(2)}$ is expressed as
\be
\l{5.4}
T_{nn}^{in,(2)}=2\pi B T^8 \Gamma (8)\int_0^1 \, dx\,x^2\sqrt{1-x^2}\,
z_{02}^{-3}\f{(|\psi^-_{02}| +x \psi^+_{02} )^3}
{(\psi^+_{02} + x |\psi^-_{02}|)^8}
\ee
For the energy-momentum tensor of gravitons which have made two bounces we have
\ba
\l{5.5}
T_{nn}^{in,(3)}
=2\pi B\int^{\infty}_0\,dE_3 \int_{-E_3}^0 dm_2 \,m_3^2\,\sqrt{E_3^2 -m_3^2}\,
z_{03}^{-3}
( E_3 \psi^-_{03} +m_3 \psi^+_{03} )^3
\exp{ \left\{- \f{E_3 \psi^+_{03} + m_3 \psi^-_{03} }{T} \right\}} \\\nonumber
=2\pi B T^8 \Gamma (8)\int_0^1 \, dx\,x^2\sqrt{1-x^2}\,
z_{03}^{-3}\f{(\psi^-_{03} - x\psi^+_{03} )^3}
{(\psi^+_{03} - x \psi^-_{03})^8},
\ea
where $z_{03}=z^{(3)}_{23}\,z^{(3)}_{12}z^{(3)}_{01}$.

Integration  over $E$ in the integrals $T^{in,(k)}_{nn}$  is performed
for $E>T_{min}$. 

Assuming that $T_{min}$ is in the region of early cosmology,
i.e. $10\lesssim \r (T_{min})/\m$,
taking  $g_*\sim 200$
\footnote{The ambiguity in $g_*$ and in $A$ in (\ref{4.91}) is due to incomplete
knowledge of the contribution of dark matter. We assume that the mass of particles
which form dark matter is in the interval $(20\div 100) GeV$. In
the period of early cosmology these particles are relativistic. Phenomenologically
the acceptable number of dark matter particles with the mass in the above interval
is $g_*\sim 100$ \cite{rub}. With the number of particle species in the
 non-supersymmetric Standard model $g_*\sim 100$, the total number is $\sim 200$.
Because of the high power of $T$  estimates weakly depend
on variations of this number.}
and using (\ref{T1}), we find that 
\be
\l{min}
T_{min}^4 \sim \f{\m}{\k^2}
.\ee
Using the relation $M^3\simeq\m M_{pl}^2$ which follows from the fit of cosmological data 
 \cite{mi1}, we have $T_{min}^4\sim (\m M_{pl})^2 /8\pi$.   
For $\m = 10^{-13} \div 10^{-9} GeV$ 
condition (\ref{min}) yields $T_{min}\sim 10^3 \div 10^5 GeV$.

 Because of the high power of $E$ in the integrals
for $T^{(k),in}$,  the main contribution to the integrals is produced 
from the region near the upper limit 
of integration $T_{max}$. Provided $T_{min}\ll T_{max}$, we set $T_{min}=0$. 
The functions $z_{k-1,k}(x)$  and the integrands 
$I^{(k)}$ for characteristic values of $x$ are given in Table 1 and Fig. 1.
For the following it is convenient to introduce the
notations
\be
\l{6.81}
T^{in,(k)}_{nn}=2\pi B T^8 \Gamma (8)\int\,dx I^{(k)} (x)
.\ee

\vspace{1cm}
\begin {table}[h]
\caption
{The functions $z^{(3)}_{k-1,k}$ for different values of $x$.}
\begin{center}
\begin{tabular}
{|c|c|c|c|c|c|c|c|}\hline
$x_3\equiv x$&.2&.3&.4&.5&.6&.7&.8\\\hline
$z_{01}^{(3)}(x )$&.870&.775&.633&.399&.112&.0034&$<10^{-4}$\\\hline
$z_{12}^{(3)}(x )$&.887&.816&.729&.617&.468&.267&.079\\\hline
$z_{23}^{(3)}(x )$&.899&.845&.787&.722&.648&.557&.438\\\hline
\end{tabular}
\end{center}
\l{Ta1}
\end{table}

\vspace{8cm}

\begin{figure}[h]
\hspace{3cm}\includegraphics[scale=.50]{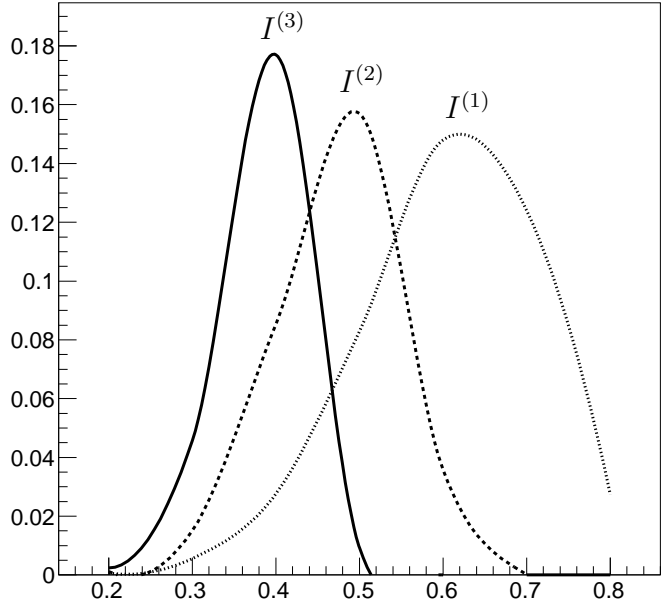}
\put(-180,230){$I^{(3)}$}
\put(-150,210){$I^{(2)}$}
\put(-110,200){$I^{(1)}$}
\caption{The integrands $I^{(k)}(x),\,\,k=1,2,3$ as the functions of $x\equiv x_3$ 
calculated for $T_R/T\gg 1$.}
\end{figure}

The integrals  $T^{in,(k)}_{nn}$ were calculated
numerically by substituting
the values for $z_{ij}(x)$.
For the integrals $\int dx I^{(k)}$ we obtain
\be
\int^1_0 dx I^{(1)}=0.0415,\qquad \int^1_0 dx I^{(2)}=0.0295,\qquad \int^1_0 dx I^{(3)}=0.0023. 
\ee 

To study dependence of $z(x)$ on $\m/\r_1$, 
we calculate $z(x)$ making use of (\ref{2.7a}) and taking into account 
the next order in $\m/\r_1$.
In Fig.2 the results for  $I^{(1)}(x)$ are compared for two values of 
$\m/\r_1= 0.1\,\,\mbox{and}\,\,\, 0.01$.
Although the values of $z(x)$ for $\m/\r_1=0.1$ and
$\m/\r_1=0.01$ considerably differ,
the values of integrals $\int\,dx I^{(1)}(\m/\r_1=0.1) =0.0460$ 
and $\int\,dx I^{(1)}(\m/\r_1=0.01) =0.0415$ are close. 

\begin{figure}[h]
\hspace{3cm}\includegraphics[scale=.45]{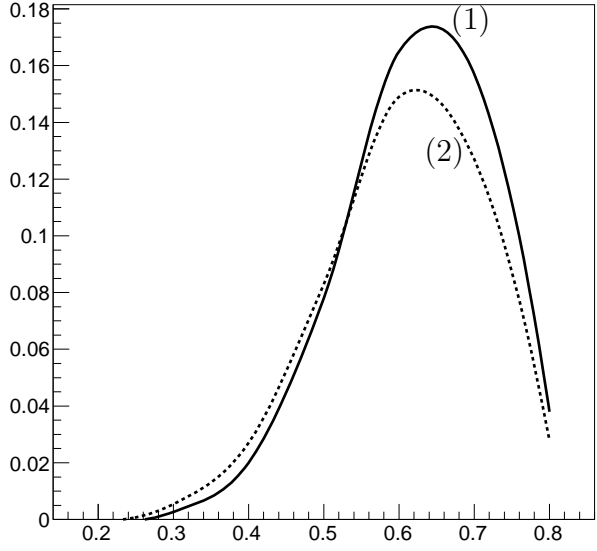}
\put(-80,210){(1)}
\put(-90,160){(2)}
\caption{Dependence of  $I^{(1)}(x)$ on  $\m/\r_1, \quad \r_1 =\r (t_1 )$, where
$t_1$ is the time of the first return of graviton to the brane calculated using (\ref{2.7b}). 
The function $I^{(1)}(x)$  calculated for $\m/\r_1 = 0.01$ (curve (1)) 
and  $\m/\r_1 = 0.1$  (curve (2)). 
For $\m/\r_1 < 0.01$ the values of $I^{(1)}(x)$ are very close to the case with $\m/\r_1 = 0$.}
\end{figure}

The limiting temperature at which the emission begins is the reheating temperature $T_R$. 
The emission energy $E_0$ is bounded by $T_R$. Using the expression of Sect.5
$E_0 =z_{0n}E (\psi^+_{0n} -x\psi^-_{0n})$,
 we have 
$$
E_{max}=T_R z_{0n}/(\psi^+_{0n} -x\psi^-_{0n} ),
$$ 
where $z_{0n}(x)$ and $\psi^\pm_{0n}=\psi^\pm (u_{0n})$ are defined
in (\ref{4.7a}) and (\ref{4.8a}) correspondingly. 
For $T_{nn}^{n,(k)}$ we obtain
\be
\l{5.11}
T_{nn}^{in,(n)}=2\pi B T^8\int_0^1 dx x^2 \sqrt{1-x^2}
 z_{0n}^{-3}(x)\f{(\psi^- (z_{0n}) -x \psi^+ (z_{0n}))^3}
{(\psi^+ (z_{0n}) -x\psi^- (z_{0n}))^8}
\g (8,  z_{0n}(x)T_R/T ),
\ee
where $\g$ is incomplete gamma-function.

If $T/T_R\lesssim 1$,
the region producing the main contribution to the integral is $1>z_{0n}>T/T_R$.
The corresponding region of $x$ is $0<x<O(1-T/T_R)$, where $1-T/T_R\ll 1$.
At small $x$ the integrand decreases as a power of $x$, and contribution from
this region is strongly suppressed.

If $T/T_R \ll 1$, the region producing the main contribution to the integral
 is  $1>z_{0n}>1-T/T_R$, where
$1-T/T_R \lesssim 1$.

\begin{figure}[h]
\includegraphics[scale=.30]{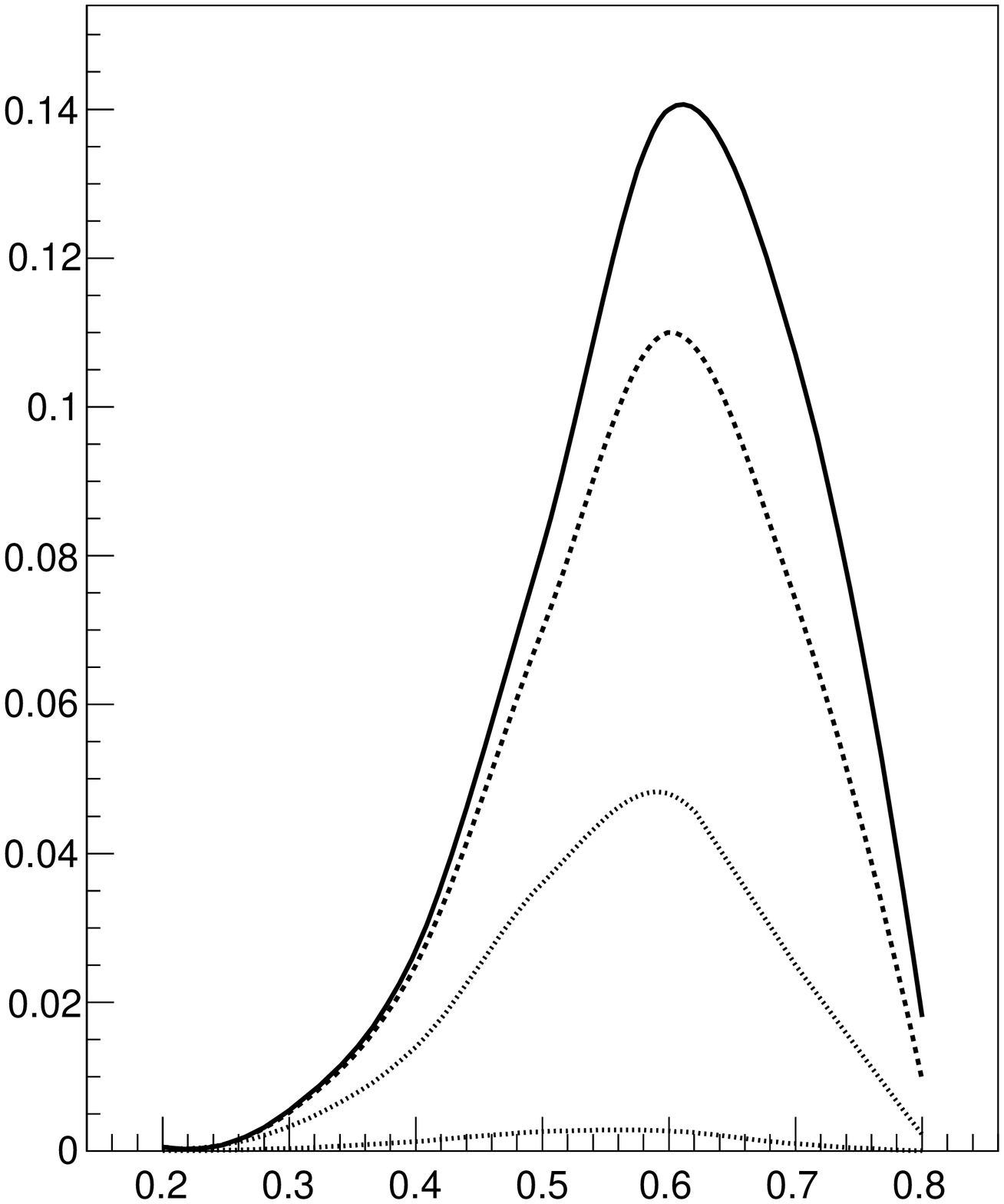}\hspace*{-1cm}
\includegraphics[scale=.30]{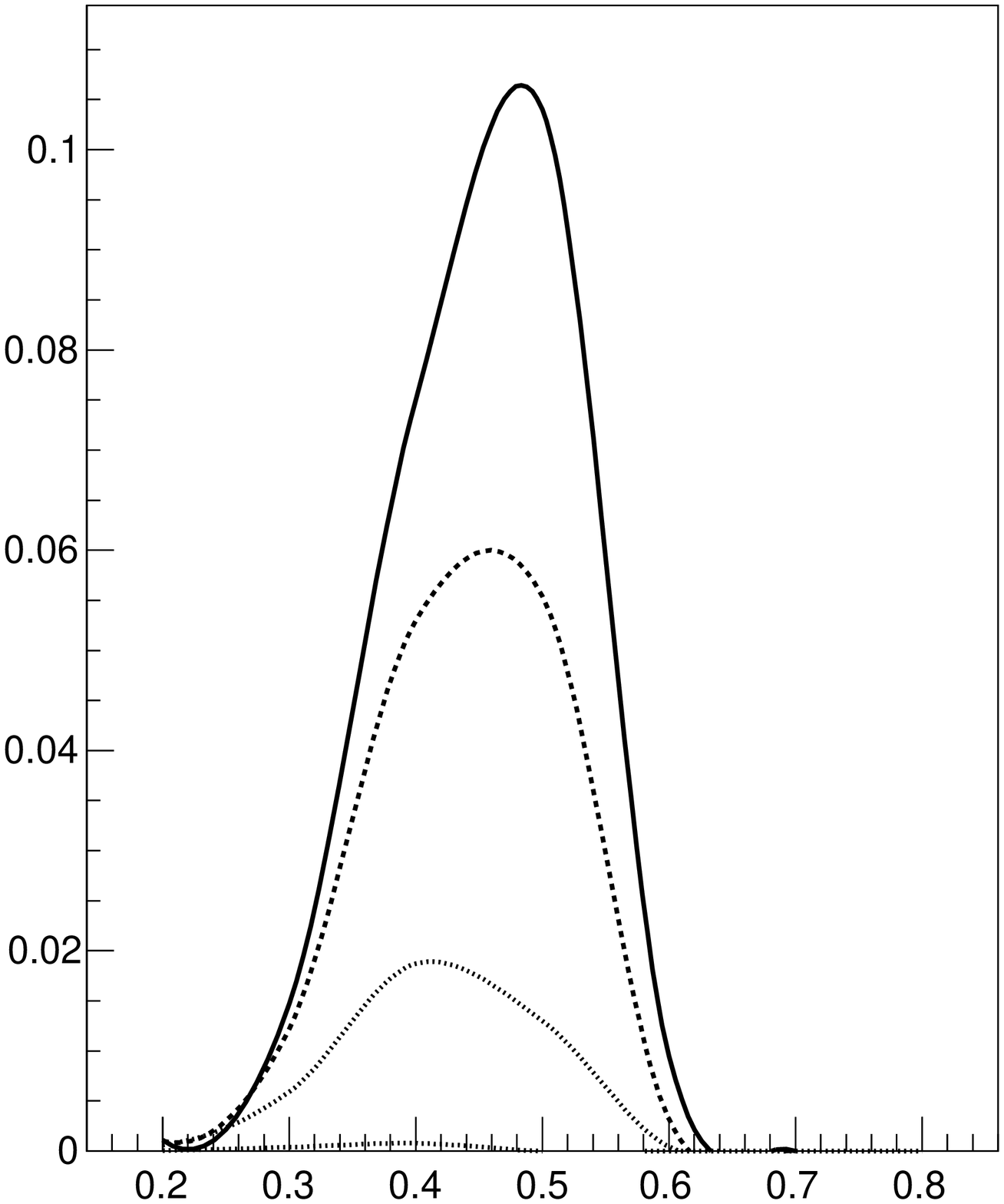}\hspace*{-1cm}
\includegraphics[scale=.30]{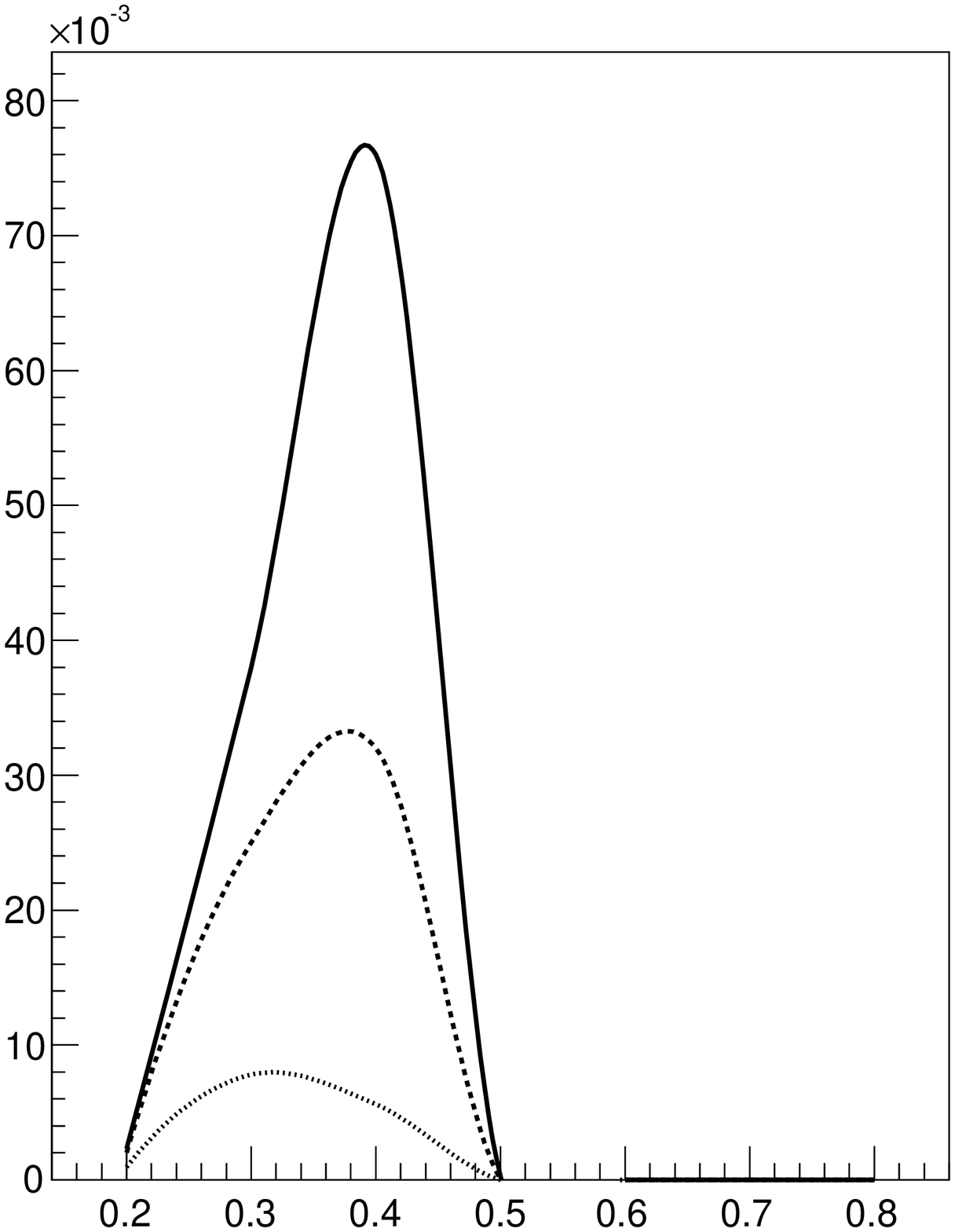}
\put(-400,150){(a)}
\put(-270,150){(b)}
\put(-140,150){(c)}
\caption{ The functions $I^{(1)}(x)\g (8, z_{01}(x)T_R/T )$ calculated  
 for $T_R/T =20,\,15,\,10,\,5$  
(curves from top to bottom) (plot (a)).  The same for 
$I^{(2)}(x)\g (8, (z_{01}z_{12})(x)T_R /T )$ (plot (b)) and 
$I^{(3)}(x)\g (8, (z_{01}z_{12}z_{23})(x)T_R /T )$ (plot (c)). 
It is seen that as $k$ increases
 the region of $x$, which produces the main contribution to $T^{in, (k)}$ is shifted to smaller
$x$ and larger $z(x)$.
}
\end{figure}

 The energy density of dark radiation satisfies the evolution  
 equation \cite{tan_him,lan2,mi3}
\be
\l{6.3}
\f{d\r_D}{dt}+4H\r_D \simeq -\f{2\r}{\m}(T^{em}_{vn} +T^{em}_{nn} -T^{in}_{nn} ), 
\ee
where $T^{in}=\sum T^{in, (k)}$ and
\ba
\l{6.4}
T^{em}_{vn}(T)=-2\pi B \,T^8 \Gamma (8)\cdot\f{\pi}{32}\f{\gamma (8,{T_R}/{T})}{\Gamma (8)}, \\
T^{em}_{nn}(T)=2\pi B\, T^8 \Gamma (8)\cdot \f{8}{105}\f{\gamma (8,{T_R}/{T}))}{\Gamma (8)}.
\ea

 
In the  period of early cosmology from the Friedmann and approximate conservation equations,
$\dot{\r} +4H\r\simeq 0$, it follows that
$$
\f{\r}{\m}= \f{1}{4\m t}=\f{2\k^2\pi^3 g_* (T)T^4}{45 (\m\,M_{pl})^2}.
$$
Here we substituted $M^3_5 \simeq \m M_{pl}^2$ \cite{mi1}. 
Eq. (\ref{6.3}) is transformed as
\be
\l{6.5}
\f{d\r_D (T)}{dT}-\f{4}{T}\r_D (T) =\f{2}{\m T}(T^{em}_{vn} +T^{em}_{nn} -T^{in}_{nn} ).
\ee
Explicitly we have
\be
\l{6.6}
\f{d\r_D}{dT}-\f{4}{T}\r_D =4\pi T^7 \Gamma (8)B \left[\left(-\f{\pi}{32}+\f{8}{105}\right)
\f{\gamma (8,{T_R}/{T})}{\Gamma (8)}-
\sum\int dx I^{(k)}(x) \f{\gamma (8, Z_{0k}{T_R}/{T}}{\Gamma (8)}\right]
\ee
Integrating (\ref{6.6}) with the boundary condition $\r_D (T_R )=0$, we obtain
\be
\r_D=-T^4\int\limits_T^{T_R} dT'\,{T'}^3\f{\Gamma (8)A\k^2}{\m 2^8 \pi^4}
\left[\left(-\f{\pi}{32}+\f{8}{105}\right)
\f{\gamma (8,{T_R}/{T'})}{\Gamma (8)}-
\sum\int dx I^{(k)}(x) \f{\gamma (8, Z_{0k}{T_R}/{T'})}{\Gamma (8)}\right]
\ee
For the ratio $\r_D/\h{\r}$ we have
\ba
\l{6.7}\nonumber
\f{\r_D (T)}{\h{\r}(T)}=\f{4725\,A}{2^5\pi^4 g_* (T) (\m M_{pl})^2} \int_{T}^{T_R }
\left[-0.224\f{\gamma (8,{T_R}/{{T'}})}{\Gamma (8)} + 
\f{32}{\pi}\sum\int dx I^{(k)}(x) \f{\gamma (8, Z_{0k}{T_R}/{{T'}})}{\Gamma (8)}\right]
T^{'3} dT'\\
=\f{4725\,A\,T^4_R}{2^5\pi^4 g_* (T) (\m M_{pl})^2 }\int_1^{T_R/T}\f{dy}{y^5}
\left[-0.224\f{\gamma (8,y)}{\Gamma (8)} + 
\f{32}{\pi}\sum\int_0^1 dx I^{(k)}(x)\f{\gamma (8,Z_{0k}y)}{\Gamma (8)}\right]\qquad{}
\ea
Here we substituted $M_5^3\simeq \m M^2_{pl}$ \cite{mi1}.

For $T_R/T > 20$ the integral is practically constant and independent of
$T$. The main contribution to the integral is produced by integration over
 the region of $y$ near the lower limit. 
Taking $T_R/T=20$ and performing integration over $y$, we obtain
\ba
\l{6.8}
0.224\int_1^{T_R/T}\f{dy}{y^5}\f{\gamma (8,y)}{\Gamma (8)}\simeq 6.56 \cdot 10^{-5}\\
\l{6.9}
\f{32}{\pi}\int_1^{T_R/T}\f{dy}{y^5}\sum_1^3\int dx I^{(k)}(x)
\f{\gamma (8,z_{0k}y)}{\Gamma (8)}\simeq 3.8\cdot 10^{-5}
.\ea
For the ratio of energy density of dark radiation to energy density of matter we have  
\be
\l{6.10}
\left|\f{\r_D}{\h{\r}}\right|\simeq 3.5\cdot 10^{-5}\f{T^4_R}{(\m M_{pl})^2}
.\ee
A typical order of constraint on magnitude of the ratio $\r_D/\h{\r}$ 
in the period of early cosmology,
which follows from primordial
nucleosynthesis, is $|\r_D/\h{\r}|\lesssim 0.07$ \cite{neutrin}.

From the gravity experiments it follows that
characteristic scale of extra dimension
$r_{extr}\sim\m^{-1}$ is less than $ 10^{-2} cm $, 
or $\m >2\cdot 10^{-12} GeV$ \cite{maart}.
For $\m \sim 10^{-12}\, GeV$
the estimate (\ref{6.10}) gives $T_R \sim  2.3\cdot 10^4\, GeV$.
This value of $T_R $ is significantly lower than usually
accepted $T_R\sim 10^5 \div 10^7\, GeV$,
indicating that large extra dimensions can  appear with
low reheating temperature.
The estimate can be improved, if  there is more complete
cancellation between two terms in (\ref{6.7}), or for larger values of $\m$.
Because of strong dependence of $\r_D/\h{\r}$ on $\m $,
the possibility of larger $\m$ seems more plausible.
For the reheating temperature $T_R\sim 10^6 GeV$ the scale of extra
dimension obtained from (\ref{6.10}) is  $\m\sim 2\cdot 10^{-9} GeV\,\,
(\m^{-1}\sim 10^{-5} cm )$. For higher reheating temperatutes the value
of $\m$ rapidly increases: $\m \sim T^2_R$.
For larger $\m$,  in the integrals for $T^{(k)}_{nn}$, the lower 
bound of integration $ E_{min}$
increases (see (\ref{min})) resulting in smaller magnitudes of the integrals $T_{nn}^{in}$.
For $T_R > 10^6 GeV$ this does not change the above results significantly,
but for smaller $T_R$ the effect of the lower bound of integration
must be taken into account.

The result of our calculations showing
that account of gravity radiation leads
to high mass scale of extra dimension can be
attributed either to an insufficient
accuracy of calculations (although our tests indicate  stability
of the result), or indicate that "too large" extra dimensions
are incompatible with this class of models.

\section{Conclusion}

In this paper in a model with extra dimension we have calculated graviton emission to the extra
dimension. Graviton emission is significant at the high-temperature period of the 
evolution of the Universe. 
The key point of the present paper is solution of  
  the system of equations for the brane trajectory and for geodesic equation for graviton trajectory.
For a given value of graviton 5-momentun  at the time of graviton detection, 
we calculated the time of graviton emission.
We obtained
the recursion relations enabling, in principle, calculate the energy-momentum tensor of
 gravitons falling back to the brane which have made an arbitrary number of bounces.  
 For the first three returns of graviton to the brane we obtained the explicit expressions for                
  the energy-momentum tensor of the gravitons falling back to the brane
and made their numerical estimates. Solving the evolution equation for the energy density of the
dark radiation, we obtained a relation connecting the reheating temperature and the scale of extra
dimension and estimated the scale of the extra dimension.

\vspace*{0.5cm}
{\large\bf Acknowledgements}

I thank S. Bunichev and  M. Smolyakov for asisstance.

This research was supported by Skobeltsyn Institute of Nuclear Physics,
Moscow State University.

\vspace*{-0.5 cm}

\end{document}